\documentclass[pdflatex,sn-mathphys]{sn-jnl}

\usepackage{textalpha}
\usepackage{lscape}
\usepackage[justification=centering]{caption}

\jyear{2021}%
\theoremstyle{thmstyleone}%
\theoremstyle{thmstyletwo}%

\theoremstyle{thmstylethree}%

\begin{document}

\title[Dangerous Events Detection]{Detection of Dangerous Events on Social Media: A Perspective Review}


\author[1]{\fnm{M. Luqman} \sur{Jamil}}\email{luqman.jamil@ubi.pt}
\equalcont{These authors contributed equally to this work.}

\author*[1,2,3]{\fnm{Sebastião} \sur{Pais}}\email{sebastiao@di.ubi.pt}
\equalcont{These authors contributed equally to this work.}

\author[1]{\fnm{João} \sur{Cordeiro}}\email{jpaulo@di.ubi.pt}
\equalcont{These authors contributed equally to this work.}

\affil*[1]{\orgdiv{Department of Computer Science}, \orgname{University of Beira Interior}, \orgaddress{\city{Covilhã}, \country{Portugal}}}

\affil[2]{\orgdiv{NOVA LINCS}, \orgname{New University of Lisboa}, \orgaddress{\city{Lisboa}, \country{Portugal}}}

\affil[3]{\orgdiv{GREYC}, \orgname{Groupe de Recherche en Informatique, Image et Instrumentation de University of Caen Normandie}, \orgaddress{\city{Caen}, \country{France}}}


\abstract{Social media is an essential gateway of information and communication for people worldwide. The amount of time spent and reliance of people on social media makes it a vital resource for detecting events happening in real life. Thousands of significant events are posted by users every hour in the form of multimedia. Some individuals and groups target the audience to promote their agenda among these users. Their cause can threaten other groups and individuals who do not share the same views or have specific differences. Any group with a definitive cause cannot survive without the support which acts as a catalyst for their agenda. A phenomenon occurs where people are fed information that motivates them to act on their behalf and carry out their agenda. One is benefit results in the loss of the others by putting their lives, assets, physical and emotional health in danger. This paper introduces a concept of dangerous events to approach this problem and their three main types based on their characteristics: action, scenarios, and sentiment-based dangerous events.}

\keywords{Dangerous Events, Social Media, Event Detection, Sentiment Analysis, Extremism, Social Network Services, Social Computing, Terrorism}



\maketitle

\section{Introduction}\label{sec1}
The extent of social media users consists of billions of people from all around the world. Initially, social media was developed with the object in mind to help connect with family and friends online. It was used to share everyday things, events, interests, and news within closed circles of family and friends. These were personal events, e.g. birthdays, weddings, vacations, graduation ceremonies, and going out. After the usability of social media was discovered, it soon caught the attention of individuals and companies that started using social media to reach more customers. Soon after, the trend became a global phenomenon where people started connecting worldwide based on their common interests. The influence of social media on people's lives and attitudes has been widely studied and established in many different perspectives \cite{intro1}, \cite{intro2}.

Although social media is a broad term, it mainly refers to Facebook, Twitter, Reddit, Instagram, and YouTube. Some social media platforms allow users to post text, photos and videos. At the same time, many other social media applications have limited options and restrictions for sharing the type of content. YouTube allows users to post videos, while Instagram only allows users to share videos and photos. 4.66 billion active internet users worldwide, and 4.2 billion users are active on social media. As of the first quarter of 2020, Facebook has 2.6 billion monthly active users globally, making it the most extensive social media network globally. Twitter is one of the leading social media with 397 million users worldwide, which is becoming increasingly prominent during events and an essential tool in politics \cite{Statista2021.}. Another study \cite{Tnewsmed} shows that Twitter is an effective and fast way of sharing news and developing stories. This trend has continued to grow since the last decade as the internet has become widespread.

However, the use of social media has become more complex in the last decade. It became a broader phenomenon because of the involvement of multiple stakeholders such as companies, groups, and other organizations. Many lobbying and public relations firms got on board and started targeting social audiences to change people's perspectives and influence their decisions. Mostly these campaigns are related to a particular individual or a company. A similar process happens in the public sphere, where people rally against or support their target. It played a significant role in different outcomes, affecting countries, people, and eventually the world. One such example is ``Arab Spring" \cite{arabspring}, which is an event that started in Tunisia and spread among other regional countries. Another example of good and bad events in political spheres of the UK and US is given in the study that uses Twitter to evaluate the perceived impact on users \cite{goodbadevents}.

The term ``event" implies typically a change which is an occurrence bounded by time and space. In the context of social media, an event can be happening on the ground or online. Different mediums can broadcast events happenings on the ground while people participate in the event through social media discussion. These kinds of events can be referred to as hybrid events. An example of such a hybrid event can be a volcano eruption where people participate in the event using online discussions on social media while it is happening on the ground. While some events solely happen online, such as gaming, marketing, and learning events. Events can be communicated in text, photos and videos across social media platforms. Many events can happen on social media platforms simultaneously, providing beneficial and prosperous information. It provides information about the event itself, but it also reveals sentiments and opinions of the general public and the direction where the events are evolving shortly. This quick interaction of users and transmission of information makes it a dynamic process that sometimes proves hard to follow the latest development, making it a challenging task.

Events also have time dimensions; it is equally important to determine the time of an event after its detection. For example, an event may have occurred in the past, happened in the present, or planned to occur in the future. Based on that, further steps would be taken accordingly as per the requirements of the situation. The events occurring on social media may directly impact the personal or social life of the man/woman. The past event can tell us people's opinions and other factors; current events can be a great source of developing a story, while future events can help us prepare in advance. The study \cite{eventtime} reviews the existing research for the detection of disaster events and classifies them in three dimensions early warning and event detection, post-disaster, and damage assessment. 

The recent example of violence in Bangladesh can explain the link between social media with real life. On Wednesday, 15 October 2021, clashes were sparked by videos and allegations that spread across social media that a Qur'an, the Muslim holy book, had been placed on the knee of a statue of the Hindu god Hanuman. The violence continued in the following days, which resulted in the deaths of 7 people, with about 150 people injured; more than 80 special shrines set up for the Hindu festival were attacked. This case shows social media's severe and robust effect on our daily lives and ground situation \cite{bngldshviolnc}. This violence was termed as ``worst communal violence in years" by New York Times. Similar episodes of violence are becoming a norm in India since the right of ring-wing politics. If there is the detection of events occurring on social media in advance, which alerts possible coming hazards, it can be countered in anticipation, significantly reducing the reaction immobilization of state forces while maximizing the protection of people at risk.

Event detection has been long addressed in the Topic Detection, and Tracking (TDT) in academia \cite{onewdet}. It mainly focuses on finding and following events in a stream of broadcast news stories shared by social media posts. Event Detection (ED) is further divided into two categories depending on the type of its task; New Event Detection (NED) and Retrospective Event Detection (RED) \cite{redev}. NED focuses on detecting a newly occurred event from online text streams, while RED aims to discover strange events from offline historical data. Often event detection is associated with identifying the first story on topics of interest through constant monitoring of social media and news streams. Other related fields of research are associated with event detection, such as; event tracking, event summarization, and event prediction. Event tracking is related to the development of some events over time. Event summarization outlines an event from the given data, while the event forecasts the next event within a current event sequence. These topics are part of the Topic Detection and Tracking (TDT) field.

Event detection is a vast research field, and there are various requirements and challenges for each task. Various terms have been used to address different events, making it complex to navigate the literature, sometimes adding confusion. We propose relevant events based on their characteristics under the umbrella term ``Dangerous Events" (DE). Section 2 defines the term ``Dangerous Events" and the main grouping criteria. In order to clarify the idea of different dangerous event types, we employ SocialNetCrawler \cite{sncrwlr} for extracting tweets that are given as examples for each type.

The rest of the paper is organized as follows: Section II presents the definition related to dangerous events in social media. Section 3 discusses event detection and related terms; Section 4 is dedicated to approaches and techniques used for event detection. Section 5 presents the warning and countering measures of dangerous events and challenges. Section 6 lists the challenges, and open problems and the paper is finally concluded in Section 7 with some interesting future research directions.

\section{Dangerous Events}\label{sec2}
According to Merriam-Webster \cite{merrwebdang}, the word ``dangerous" means involving possible injury, pain, harm, or loss characterized by danger. In that context, we define a dangerous event as the event that poses any danger to an individual, group, or society. This danger can come in many shapes and intensities. The objective is to draw a fine line between normal, harmless, unpleasant, and extreme, abnormal and harmful events. Less sensitive, unpleasant, and disliked events do not compel the person to feel threatened. While, in the case of dangerous events, the person will feel fearful, unsafe, and threatened. This provides the objective to approach the term ``event" in a broader sense to address the common element of all such events. The details of dangers can always be discussed in detail, providing the necessity of the situation; for example, a natural disaster proceeds urgent hate speech. In other words, the first one requires an immediate response with no time to lose, while the latter can allow some time to take action. 

Dangerous events can be anomalies, novelty, outliers, and extreme. These terms can be used to refer to positive or negative meanings. However, Not all anomalies, novelties, and extremes are dangerous, but all dangerous events fulfil one or all of those conditions (extreme, anomaly, novelty). Authors in \cite{extremsentilax} proposed an unsupervised approach to detect extreme sentiments on social media. It shows that Positive Extreme sentiments can be detected and differentiated from everyday positive sentiments. Therefore, it may be concluded that extreme negative sentiments are likely to turn into dangerous events.

Grouping and defining dangerous events based on their characteristics is another challenging task, and it can help address the issue of approaching different types of dangerous events by narrowing it down to specific details. We will define three broad categories of dangerous events with commonality among them.
\begin{enumerate}
\item Scenario-based Dangerous Events
\item Sentiment-based Dangerous Events
\item Action-based Dangerous Events
\end{enumerate}

\begin{figure}[h]%
\centering
\includegraphics[width=0.9\textwidth]{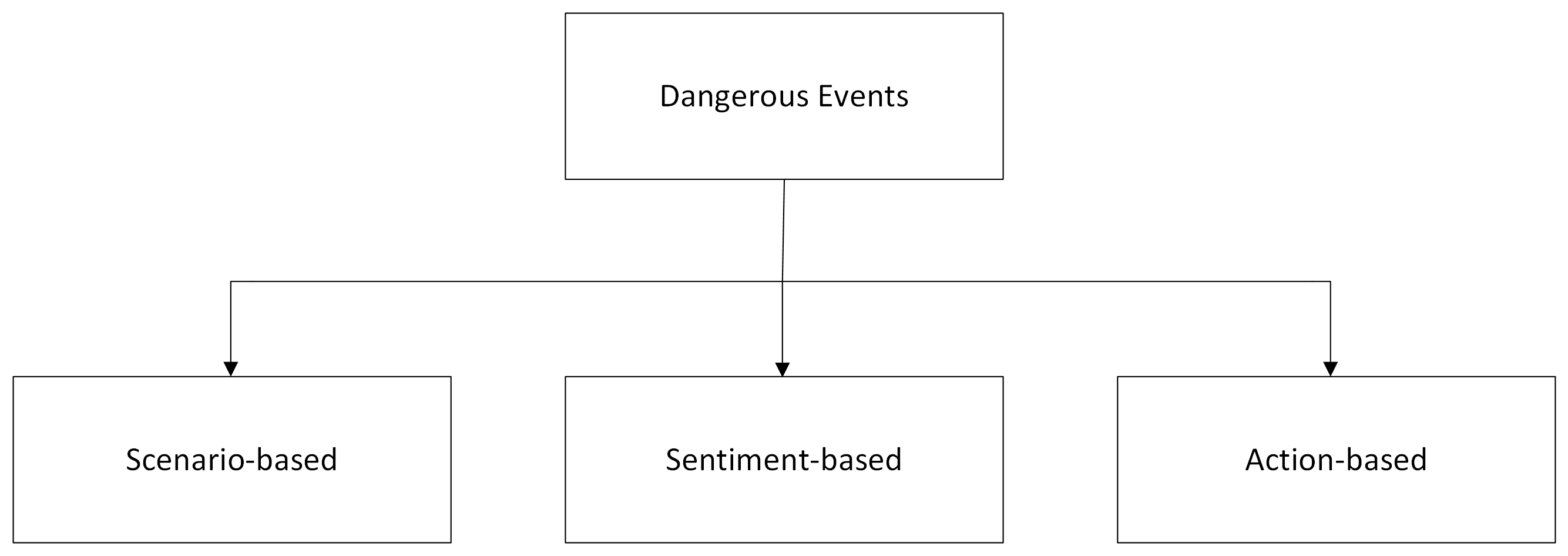}
\caption{Dangerous Events and it's categories.}\label{figde}
\end{figure}

Figure \ref{figde} gives the depiction of dangerous events and their categories. In the following subsections, we will outline the definition for each type of dangerous event.

\subsection{Scenario-based Dangerous Events}
We refer to the word ``scenario" as the development of events. These events are unplanned and unscripted, and most of the time, they occur naturally. Some planned events can also turn into surprising scenarios. For example, a peaceful protest can turn into a riot, like in 2020 where a peaceful protest against corona restrictions in Germany turned into an ugly situation when the rally was hijacked by right-wing extremists, which ended up storming Parliament building and exhibiting right-wing symbols and slogans \cite{germanparl}. 

Detecting and tracking natural disasters on social media have been investigated intensively, and studies \cite{sc-based1} have proposed different methods to identify those disasters by various means. The aim of these studies has been mainly to tap into the potential of social media to get the latest updated information provided by social media users in real-time and identify the areas where assistance is required. These categories are considered scenario-based dangerous events in this paper, including earthquakes, force majeure, hurricanes, floods, tornadoes, volcano eruptions, and tsunamis. Although each calamity's nature is different, the role of social media in such events provides a joint base to approach them as scenario-based dangerous events. A supposed example of scenario-based danger is obtained using the crawler tool SocialNetCrawler, which can be accessed using the link\footnote{\url{http://sncrawler.di.ubi.pt/}}:

\textit{``@politicususa BREAKING: 
Scientists predict a tsunami will hit 
Washington, DC on 1/18/2020
We Are Marching in DC… https://t.co/3af4ZhyV3J"}

\subsection{Sentiment-based Dangerous Events}
Sentiment Analysis (SA), also known as Opinion Mining (OM), is the process of extracting people's opinions, feelings, attitudes, and perceptions on different topics, products, and services. The sentiment analysis task can be viewed as a text classification problem as the process involves several operations that ultimately classify whether a particular text expresses positive or negative sentiment \cite{sncamb}. For example, A micro-blogging website like Twitter is beneficial for predicting the index of emerging epidemics. These are platforms where users can share their feelings which can be processed to generate vital information related to many areas such as healthcare, elections, reviews, illnesses, and others. Previous research suggests that understanding user behaviour, especially regarding the feelings expressed during elections, can indicate the outcome of elections \cite{saelctn}. 

Sentiments can be positive and negative, but for defining sentiment-based dangerous events, the applicable sentiments are negatives and, in some instances, negative extremes. Online radicalization can be attributed to this threat related to extreme negative sentiments towards certain people, countries, and governments. Such extreme negative sentiments can result in protests, online abuse, and social unrest. Detecting these events can help reduce its impact by allowing the concerned parties to counter beforehand. A hypothetical example of sentiment-based dangerous example of a tweet obtained using SocialNetCrawler is given below:

\textit{``RT @Lrihendry: When Trump is elected in 2020, I’m outta here. 
It’s a hate-filled sewer. 
It is nearly impossible to watch the hateful at…"}

\subsection{Action-based Dangerous Events}
The action involves human indulgence in an event. Various types of actions happen on the ground that can be detected using social media. Actions can be of many types, but we point out actions that are causing harm, loss, or threat to any entity, which again shares the common attribute of negativity and is highly similar to previously defined types of dangerous events. Some examples of Action-based dangerous events can be prison breaks, terrorist attacks, military conflicts, shootings, etc. Several studies have been published focusing on one or more types of such actions-based events. The study \cite{antifas} focuses on anti-fascist accounts on Twitter to detect acts of violence, vandalism, de-platforming, and harassment of political speakers by Antifa. An assumed example of action-based example acquired using SocialNetCrawler is given below:

\textit{``RT @KaitMarieox: This deranged leftist and LGBT activist named Keaton Hill assaulted and threatened to kill @FJtheDeuce, a black conservati…''}

\section{Event Detection methods and techniques}

Event Detection has been a popular topic in the research community. Several methods and techniques have been proposed to detect events depending on different requirements. These methods directly depend on the type of task and the data available. As such, they were detecting events from image data is undoubtedly different from text data. However, in this work, we only refer to event detection techniques related to text data, particularly data obtained from social media platforms. 

\begin{figure}[H]
\centering
\includegraphics[width=0.80\textwidth]{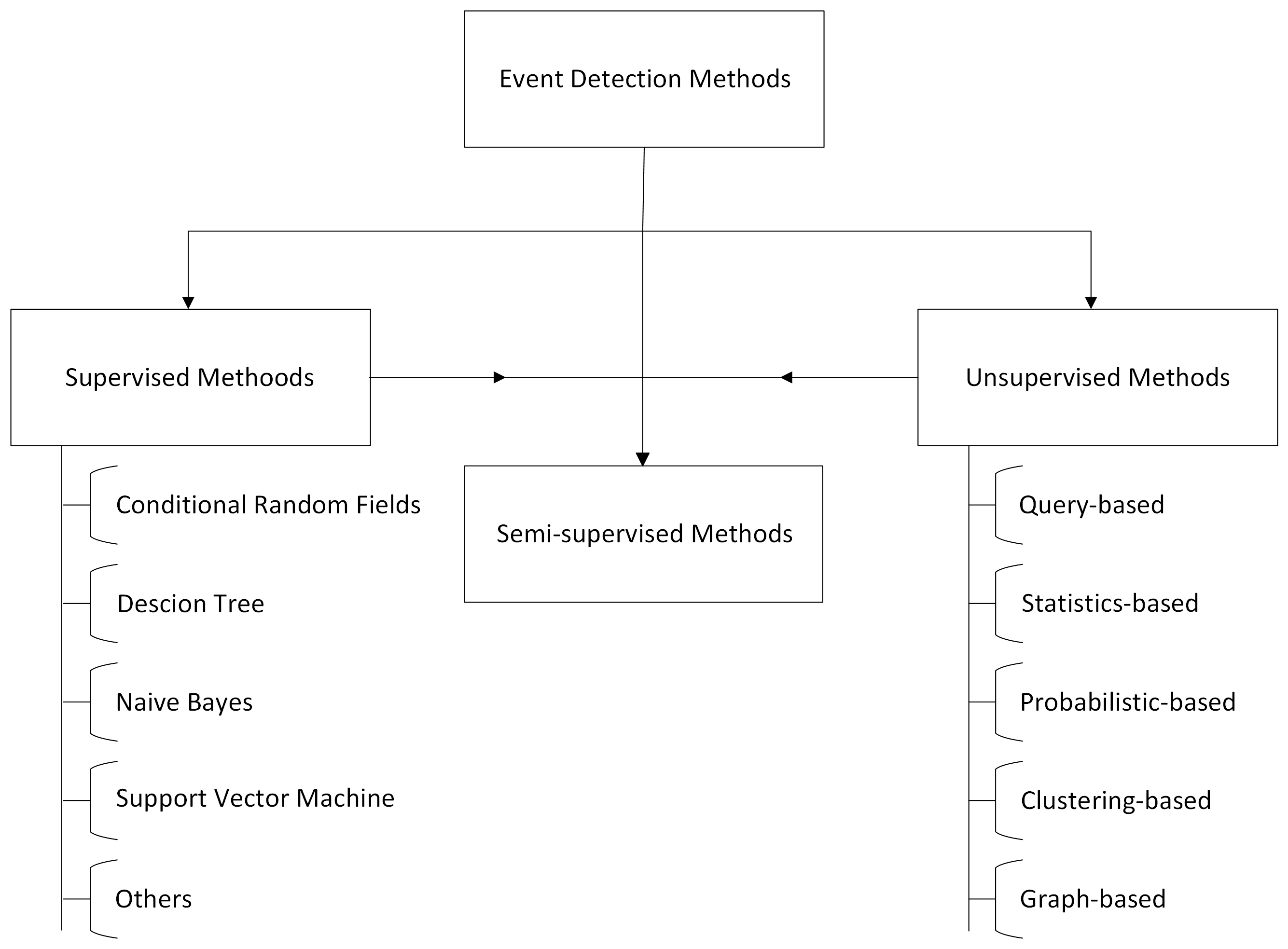}
\caption{Classification of ED methods.} \label{evmethds}
\end{figure}

Event detection methods and techniques revolve around a few basic approaches. Two approaches that are being used in event detection are document-pivot and feature-pivot. What differs in these approaches is mainly the clustering approach, the way documents are used to feature vectors, and the similarity metric used to identify if the two documents represent the same event or not. Another approach is the topic modelling approach, primarily based on probabilistic models.

\begin{figure}[H]
\centering
\includegraphics[width=0.8\textwidth]{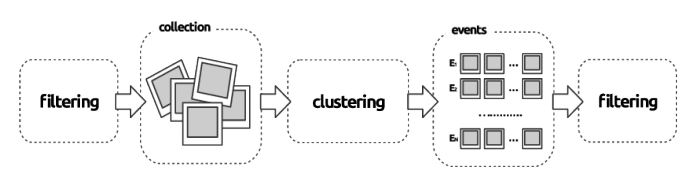}
\caption{Event Detection using Document-pivot approach \cite{31}. \label{doc:fig} }
\end{figure}

It originates from the Topic Detection and Tracking task (TDT) field and can be seen as a clustering issue. \textbf{Document-pivot approach} detects events by clustering documents based on document similarity as given in Figure \ref{doc:fig}. Documents are compared using cosine similarity with Tf-IDF (term frequency-inverse document frequency) representations, while a Locality Sensitive Hashing (LSH) \cite{lcltyhashing} scheme is utilized to retrieve the best match rapidly. 

\begin{figure}[H]
\centering
\includegraphics[width=0.8\textwidth]{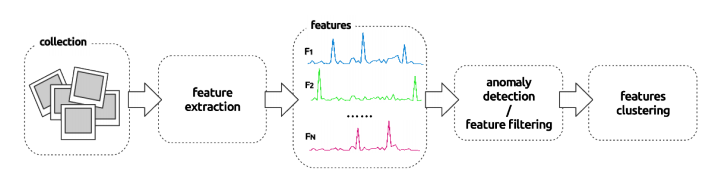}
\caption{Event Detection using Feature-pivot approach \cite{31}. \label{featurebsd}}
\end{figure}

This technique was initially proposed for the analysis of timestamped document streams. The bursty activity is considered an event that makes some of the text features more prominent. The features can be keywords, entities and phrases. \textbf{Feature-pivot Approach} clusters together with terms based on the pattern they occur as shown in the Figure \ref{featurebsd}. A study \cite{featurenaive} uses a Naive Bayes classifier to learn the selected features such as keywords to identify civil unrest and protests and accordingly predict the event days.

\begin{figure}[H]
\centering
\includegraphics[width=0.4\textwidth]{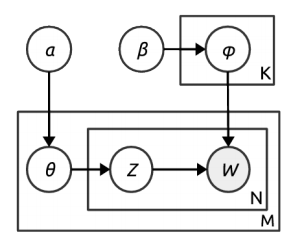}
\caption{LDA - A common topic modeling technique \cite{31}. } \label{ldafig}
\end{figure}

\textbf{Topic modelling approaches} are based on probabilistic models which detect events in social media documents in a similar way that topic models identify
latent topics in text documents. In the beginning, topic models depended on word occurrence, where the text corpora were given as a mixture of words with latent model topics and the set of identified topics were given as documents. Latent Dirichlet Allocation (LDA) \cite{jelodar2019latent} is the most known probabilistic topic modelling technique. It is a hierarchical Bayesian model where a topic distribution is supposed to have a sparse Dirichlet prior. The model is shown in the Figure. \ref{ldafig}, where α is the parameter of the Dirichlet before the per-document topic distribution θ and φ is the word distribution for a topic. K represents the number of topics, M represents the document number, while N gives the number of words in a document. If the word W is the only observable variable, the learning of topics, word probabilities per topic, and the topic mixture of each document is tackled as a problem of Bayesian inference solved by Gibbs sampling.

Many methods are proposed for the detection of events. This event detection (ED) methods are mainly categorized under the category of supervised and unsupervised, as shown in Figure \ref{evmethds}. Supervised methods include support vector machine (SVM), Conditional random field (CRF), Decision tree (DT), Naive Bayes (NB) and others. At the same time, the unsupervised approaches include query-based, statistical-based, probabilistic based, clustering-based, and graph-based.

\subsection{Event Detection Datasets}
Due to the growth of the internet and related technologies, the research in event detection has experienced significant interest and effort. However, the benchmark datasets for event detection witnessed slow progress. This can be attributed to the complexity and costliness of annotating events that require human input. There are a handful number of datasets available that covers event detection. These datasets are mostly limited to the small size of data and very restricted types of events. They address specific domains based on certain features. This also raises issues while using a data-hungry deep learning model and typically requires balanced data for each class. Some of these datasets are briefed in the following paragraphs. Table \ref{tab:edds} gives the comparison of the discussed datasets and knowledge bases.

MAVEN \cite{wang2020MAVEN} which stands for MAssive eVENt detection dataset, offers a general domain event detection dataset manually annotated by humans. It uses English Wikipedia and FrameNet (Baker et al., 1998) documents for building the dataset. It contains 111,611 various events and 118,732 events mentioned. The authors claim this to be the largest available human-annotated event detection dataset. There are 164 different types of events, representing a much wider range of public domain events. The event types are grouped under five top-level types: action, change, scenario, sentiment, and possession. 

EventWiki \cite{ge2018eventwiki} is a knowledge base of events, which consists of 21,275 events containing 95 types of significant events collected from Wikipedia. EventWiki gives four kinds of information: event type,
event info-box, event summary, and full-text description. Authors claim to be the first knowledge base of significant events, whereas most knowledge bases focus on static entities such as people, locations, and organizations.

The EventKG \cite{Abdollahi2020EventKGClickAD} is a multilingual1 resource incorporating event-centric information extracted from several large-scale knowledge graphs such as Wikidata, DBpedia and YAGO, as well as less structured sources such as the Wikipedia Current Events Portal and Wikipedia event lists in 15 languages. It contains
details of more than 1,200,000 events in nine languages. Supported languages include; English, French, German, Italian, Russian, Portuguese, Spanish, Dutch, Polish, Norwegian, Romanian, Croatian, Slovene, Bulgarian and Danish.

EVIN \cite{EVIN} which stands for EVents In News, describes a method that can extract events from a news corpus and organize them in relevant classes. It contains 453 classes of event types and 24,348 events extracted from f 300,000 heterogeneous news articles. The news articles used in this work are from a highly diverse set of newspapers and other online news providers (e.g., http://aljazeera.net/,
http://www.independent.co.uk, http://www.irishtimes.com,
etc.). These news articles were crawled from the external links mentioned on Wikipedia pages while ignoring the content of Wikipedia pages to get the articles from the original website source.

\begin{table}[h!]
  \begin{center}
    \resizebox{\textwidth}{!}{%
    \begin{tabular}{lllllll}
    \toprule 
        \textbf{Dataset} & \textbf{Events} & \textbf{Event types} & \textbf{Document Source}& \textbf{Language}& \textbf{Year} &\textbf{Reference}  \\
         \midrule 
            MAVEN & 111, 611 & 164 & English Wikipedia \& FrameNet  & English & 2020 & \cite{wang2020MAVEN} \\
            EventWiki & 21,275 & 94 & English WIkipedia & English& 2018 & \cite{ge2018eventwiki} \\
            EventKG & 1,200,000 & undefined & DBpedia \& YAGO... & Multilingual(9) & 2020 & \cite{Abdollahi2020EventKGClickAD} \\
            EVIN &  24,348 & 453
 & news corpus &English &2014 & \cite{EVIN}\\
        \bottomrule 
    \end{tabular}} \caption{Comparison of Related Event Detection Datasets }     \label{tab:edds}
  \end{center}
\end{table}

\subsection{Supervised Methods}
Supervised methods are expensive and lengthy as they require labels and training, and this becomes difficult for more extensive data where the cost of training the model is higher and time-consuming. Some of the supervised methods for event detection are discussed below. 

\textbf{Support Vector Machines (SVM):}
Support vector machines are based on the principle of minimizing structural risks \cite{statlrnth} of computer learning theory. Minimizing structural risks
is to find an assumption h for which we can guarantee the lowest true error. The real error in h is the probability that h will make an error in a sample test selected at random. An upper limit can be used to connect the true error of a hypothesis h with the error of h in the training set and the complexity of H (measured by VC-Dimension), the space of hypotheses which contains h \cite{statlrnth}. The supporting vector machines find the hypothesis h, which (approximately) minimize this limit on the true error by controlling effectively and efficiently the VC dimension of H.\cite{txtcatsvm}

It has been confirmed in many works that SVM is one of the most efficient algorithms for text classification. The accuracy of 87\% was achieved to classify the traffic or nontraffic events on Twitter. It was able to identify valuable information regarding traffic events through Twitter \cite{incdetsm}. SVM
combination with incremental clustering technique was applied to detect social and real-world events from photos posted on Flicker site \cite{smedgrphmdl}.\\

\textbf{Conditional Random Fields (CRF):}The CRFs is an essential type of machine learning model developed based on the Maximum Entropy Markov Model (MEMM). It was first proposed by Lafferty et al. (2001) as probabilistic models to segment and label sequence data, inherit the advantages of the previous models, increase their efficiency, overcome their defects, and solve more practical problems \cite{crfprbmdl}.
A conditional Random Field (CRF) classifier was learned to extract the artist name and location of music events from a corpus of tweets \cite{edinsmfeeds}.

\textbf{Decision Tree (DT):}
Decision tree learning is a supervised machine learning technique for producing a decision tree from training data. A decision tree is also referred to as a classification tree or a reduction tree, and it is a predictive model which draws from observations about an item to conclusions about its target value. In the tree structure, leaves represent classifications (also referred to as labels), non-leaf nodes are features, and branches represent conjunctions of features that lead to the classification \cite{artanlyzsft}.
A decision tree classifier called gradient boosted was used to anticipate whether the tweets consist of an event concerning the target entity or not.

\textbf{Naïve Bayes (NB):}
Naïve Bayes is a simple learning algorithm that uses the Bayes rule and a strong assumption that the attributes are conditionally independent if the class is given. Although this independence assumption is often violated in practice, naïve Bayes often provides competitive accuracy. Its computational efficiency and many other distinctive features result in naïve Bayes being extensively applied in practice.

Naïve Bayes gives a procedure for using the information in sample data to determine the posterior probability P(y\textbar x) of each class y, given an object x. Once we have such estimates, they can be then used for classification or other decision support applications. \cite{enclypmlnaivebyes}

\subsection{Unsupervised Methods}
The unsupervised method does not usually require training or target labels. However, they can depend on specific rules based on the model and requirements. The unsupervised methods being used for event detection are discussed below. Many unsupervised methods are developed by scientists who are grouped into different categories that are described in the following subsections. 

\textbf{Query Based Methods:}
Query-based methods are based on queries and simple rules to identify planned rules from multiple websites. e.g., YouTube, Flicker, Twitter. An event's temporal and spatial information was extracted and then used to inquire about other social media websites to obtain relevant information.\cite{qureybsd} 
The query-based method requires predefined keywords if there are many keywords to avoid unimportant events.

\textbf{Statistical Based Methods:}
Different researchers under this category introduced many methods. 
For example, the average frequency of unigrams was calculated to find the significant unigrams (keywords) and combine those unigrams to illustrate the trending events. \cite{ unsuptopkextrct} The attempt was made to detect the hot events by identifying burst features (i.e., unigram) during different time windows. Each unigram bursty feature signal was then converted into a frequency domain. They were using Discrete Fourier Transformation (DFT). However, DFT was not able to detect the period when there is a burst which is very important in ED process\cite{twitternewsrepo}. 

\textbf{Wavelet Transformation(WT):} Another technique called Wavelet Transformation (WT) was introduced to assign signals to each unigram feature. WT technique is different from DFT in term of isolating time and frequency and provide better results\cite{edintwitter}. 
A new framework was proposed that integrated different unsupervised techniques. For example, LDA, NER, bipartite graph clustering algorithm based on relation and centrality scores to discover hidden events and extract their essential information such as time, location, and people that have been involved \cite{edsocialweb}.

\textbf{Named Entity Relation(NER):} Named Entity Relation (NER) identifies increasing weights for the proper noun features. A proposed technique applied tweet segmentation to get the sentences containing more phrasing words instead of unigrams. Later, they computed the TFIDF of these sentences and user frequency and increased weights for the proper noun features identified by Named Entity Relation (NER). Li et al. (2012a)first applied tweets classified them using K-Nearest Neighbor (KNN) to identify the events from tweets published by Singapore users\cite{twitterbsded}.

Weiler et al. (2014) \cite{eventidentity} used shifts of terms computed by Inverse Document Frequency (IDF) over a simple sliding window model to detect events and trace their evolution. Petrović et al. ( 2010) \cite{fsdtwitter} modified and used Locality Sensitive Hashing (LSH) to perform First Story Detection (FSD) task on Twitter.

\begin{figure}[H]
\centering
\includegraphics[width=0.8\textwidth]{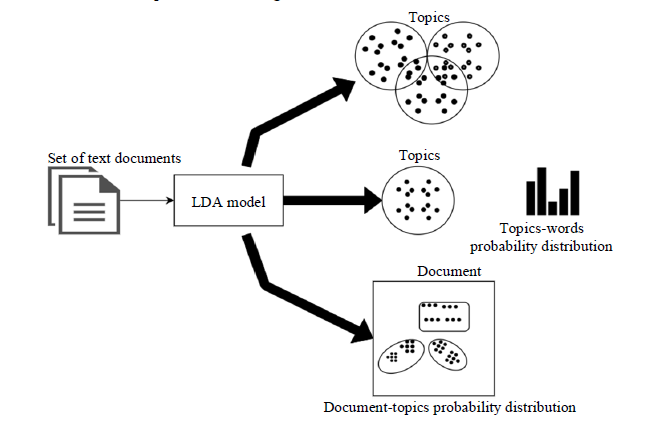}
\caption{Topic Modeling in LDA \cite{srvyedtxtstrm}. } \label{lda}
\end{figure}

\textbf{Probabilistic Based Methods:}
Latent Dirichlet Allocation (LDA) and Probabilistic Latent Semantic Indexing (PLSI) are topic modelling methods that are being used for event detection. In LDA, each document has many topics, and it is supposed to have a group of topics for each document. The model is shown in Figure \ref{lda}.

LDA worked well with news articles and academic abstracts, but it fell short for small texts. However, the LDA model has been improved by adding tweet pooling schemes and automatic labelling. Pooling schemes include basic scheme, author scheme, burst term scheme, temporal scheme, and hashtag scheme tweets published under the same hashtag. The experiment results proved that the hashtag scheme produced the best clusters results \cite{ldatpc}. However, LDA defines the number of topics and terms per topic in advance, inefficiently implementing it over social media. 
 
\textbf{Clustering-Based Method:}
Clustering-based methods mainly rely on selecting the most informative features, which contribute to event detection, unlike supervised methods, which need labelled data for prediction. It contributes to detecting events more accurately.

\begin{figure}[H]
\centering
\includegraphics[width=0.8\textwidth]{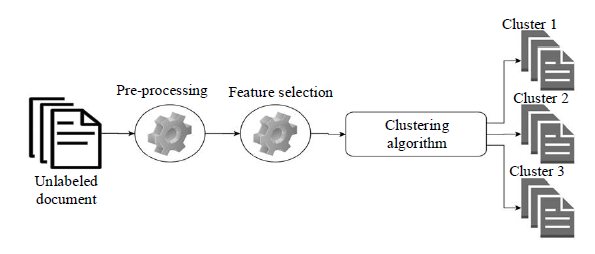}
\caption{Clustering-Based method \cite{srvyedtxtstrm}. } 
\end{figure}

Many clustering-based methods exist for text data, and K-means is a famous clustering algorithm. A novel dual-level clustering was proposed to detect events based on news representation with time2vec \cite{dualclstr}. Clustering-based methods have been employed in various ways and other techniques such as NER, TFIDF and others in different tasks, but the ideal clustering technique is still yet to come.

\textbf{Graph-Based Methods:}
Graph-based methods consist of nodes/vertices representing entities and edges representing the relationship between the nodes. Valuable information can be extracted from these graphs by grouping a set of nodes based on the set of edges. Each generated group is called a cluster/graph structure, and it is also known as community, cluster or module. The links between different nodes are called intra-edges. Meanwhile, links that connect different communities are called inter-edges.

\begin{figure}[H]
\centering
\includegraphics[width=0.8\textwidth]{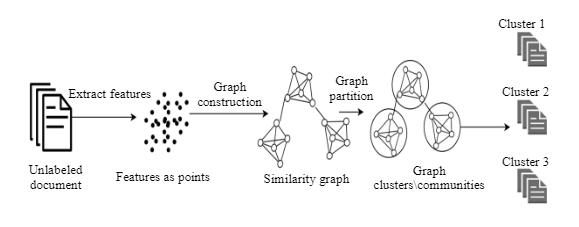}
\caption{Graph-based clustering method \cite{srvyedtxtstrm}. } \end{figure}

\subsection{Semi-Supervised Methods}
Semi-supervised learning combines both supervised and unsupervised learning methods. Typically, a small number of labelled and largely unlabeled data is used for training purposes. Sometimes they are also referred to as the hybrid method. If there is a vast number of unlabeled data combined with insufficient labelled data, it can affect the classification accuracy. It is also referred to as imbalanced training data.

Similarly, if there is no labelled data for a particular class, the classification can become inefficient and accurate. Some of the semi-supervised methods include self-training, generative models and graph-based methods. A semi-supervised algorithm based on tolerance roughest and ensemble learning is recommended for such kinds of problems \cite{ensamblern}. The missing class is extracted by approximation from the dataset and used as the labelled sample. The ensemble classifier iteratively builds the margin between positive and negative classes to estimate negative data further since negative data is mixed with the positive data. Therefore, classification is done without training samples by applying a hybrid approach, and it saves the cost of getting labelled data manually, especially for larger datasets.

\section{Discussion}
This section discusses different works related to event detection that are categorized under the types proposed earlier in this work. The types of events are scenario-based, sentiment-based, and action-based dangerous events. Each work is described in this section and its event type and technique. Furthermore, this section also discusses the research related to event prediction. Table \ref{detable} illustrates  different type of events detection from social media.
\subsection{Detection of Different/Dangerous Events on Social Media}
Nourbakhsh et al. \cite{nourbakhsh2017breaking} address natural and artificial disasters on social media. They identified events from local news sources that may become global breaking news within the next 24 hours. They used Reuters News Tracer, a real-time news detection and verification engine. It uses a fixed sphere decoding (FSD) algorithm to detect breaking stories in real-time from Twitter. Each event is shown as a cluster of tweets engaging with that story. By considering different data features, they applied SGD and SVM classifier that detects breaking disasters from postings of local authorities and local news outlets.

Sakaki et al. \cite{equake2010} leverage Twitter for detecting earthquake occurrence promptly. They propose a method to scrutinize the real-time interaction of earthquakes events and, similar to detect a target event. Semantic analyses were deployed on tweets to classify them into positive and negative classes. The target for classification is two keywords; earthquake or shaking, which are also addressed as query words. Total of 597 positive samples of tweets that report earthquake occurrence is used as training data. They also implemented filtering methods to identify the location, and an application called the earthquake reporting system in Japan.

Liu et al. \cite{crisisbert} aims for crisis events. They propose a state-of-the-art attention-based deep neural networks model called CrisisBERT to embed and classify crisis events. It consists of two phases which are crisis detection and crisis recognition. In addition, another model for embedding tweets is also introduced. The experiments are conducted on C6 and C36 datasets. According to the authors, these models surpass state-of-the-art performance for detection and recognition problems by up to 8.2\% and 25.0\%, respectively.

Archie et al. \cite{hurrearth} proposed an unsupervised approach for the detection of sub-events in major natural disasters. Firstly, noun-verb pairs and phrases are extracted from tweets as an important sub-event prospect. In the next stage, the semantic embedding of extracted noun-verb pairs and phrases is calculated and then ranked against a crisis-specific ontology called management of Crisis (MOAC) ontology. After filtering these obtained candidate sub-events, clusters are formed, and top-ranked clusters describe the highly important sub-events. The experiments are conducted on Hurricane Harvey and the 2015 Nepal Earthquake datasets. According to the authors, the approach outperforms the current state-of-the-art sub-event identification from social media data.

Forests fire have become a global phenomenon due to rising droughts and increasing temperatures which is often attributed to global warming and climate change. The work \cite{firehaze} tests the usefulness of social media to support disaster management. However, the primary data for dealing with such incidents come from NASA satellite imagery. The authors use GPS-stamped tweets posted during 2014 from Sumatra Island, Indonesia, which experiences many haze events. Twitter has proven to be a valuable resource during such events, confirmed by performing analysis on the dataset. Furthermore, the authors also announced the development of a tool for disaster management. 

Huang et al. \cite{emergemcyweibo} focuses on emergency events. They consider the various type of events under the term ``emergency events". It includes infectious disease, explosions, typhoons, hurricanes, earthquakes, floods], tsunamis, wildfires, and nuclear disasters. The model must automatically identify the attribute information 3W (What, When, and Where) of emergency events to respond in time. Their proposed solution contains three phases, the classification phase, the extraction phase, and the clustering phase, and it is based on the Similarity-Based Emergency Event Detection (SBEED) framework. The experiment is done using the Weibo dataset. Different classification models such as KNN, Decision Trees, Naïve Bayes, Linear SVC (RBF), and Text-CNN are used in the classification phase. Secondly, time and location are extracted from the classification obtained. Lastly, an unsupervised dynamical text clustering algorithm is deployed to cluster events depending on the text-similarity of type, time and location information. The authors claim superiority of the proposed framework having good performance and high timeliness that can be described what emergency, and when and where it happened.
  
Pais et al. present an unsupervised approach to detect extreme sentiments on social networks. Online wings of radical groups use social media to study human sentiments engaging with uncensored content to recruit them. They use people who show sympathy for their cause to further promote their radical and extreme ideology. The authors developed a prototype system composed of two components, i.e., Extreme Sentiment Generator (ESG) and Extreme Sentiment Classifier (ESC). ESG is a statistical method used to generate a standard lexical resource called ExtremesentiLex that contains only extreme positive and negative terms. This lexicon is then embedded to ESC and tested on five different datasets. ESC finds posts with extremely negative and positive sentiments in these datasets. The result verifies that the posts that were previously classified as negatives or positives are, in fact, extremely negatives or positives in most cases.
 
COVID-19 pandemic has forced people to change their lifestyles. Lockdown further pushed people to use social media to express their opinions and feelings. It provides a good source for studying users' topics, emotions, and attitudes discussed during the pandemic. The authors of work \cite{covsent} collected two massive COVID-19 datasets from Twitter and Instagram. They explore data with different aspects, including sentiment analysis, topics detection, emotions, and geo-temporal. Topic modelling on these datasets with distinct sentiment types (negative, neutral, positive) shows spikes on specific periods. Sentiment analysis detects spikes on specific periods and identifies what topics led to those spikes attributed to economy, politics, health, social, and tourism. Results showed that COVID-19 affected significant countries and experienced a shift in public opinion. Much of their attention was towards China. This study can be very beneficial to read people's behaviour as an aftermath; Chinese people living in those countries also faced discrimination and even violence because of the Covid-19 linked with China.
 
Plaza-del-Arco et al. \cite{hateoffense} investigate the link of hate speech and offensive language(HOF) with relevant concepts. Hate speech targets a person or group as a negative opinion, and it is related to sentiment analysis and emotion analysis as it causes anger and fear inside the person experiencing it. The approach consists of three phases and is based on multi-task learning (MTL). The setup is based on BERT, a transformer-based encoder pre-trained on a large English corpus. Four sequence classification heads are added to the encoder, and the model is fine-tuned for multi-class classification tasks. The sentiment classification task categorizes tweets into positive and negative categories, while emotion classification classifies tweets into different emotion categories (anger, disgust, fear, joy, sadness, surprise, enthusiasm, fun, hate, neutral, love, boredom, relief, none). The offence target is categorized as an individual, group, and unmentioned to others. Final classification detects HOF and classifies tweets into HOF and non-HOF. 
 
Kong et al. \cite{farrightextreme} explore a method that explains how extreme views creep into online posts. Qualitative analysis is applied to make ontology using Wikibase. It proceeded from the vocabulary of annotations such as the opinions expressed in topics and labelled data collected from three online social networking platforms (Facebook, Twitter, and Youtube). In the next stage, a dataset was created using keyword search. The labelled dataset is then expanded to using a looped machine learning algorithm. Two detailed case studies are outlined with observations of problematic online speech evolving the Australian far-right Facebook group. Using our quantitative approach, we analyzed how problematic opinions emerge. The approach exhibits how problematic opinions appear over time and how they coincide.
 
Demszky et al.\cite{plrztnpolitk} highlights four linguistic dimensions of political polarization in social media, which includes; topic choice, framing, affect and apparent force. These features are quantified with existing lexical methods. Clustering of tweet embeddings is proposed to identify important topics for analysis in such events. The method is deployed on 4.4M tweets related to 21 mass shootings. Evidence proves the discussions on these events are highly polarized politically, and it is driven by the framing of biased differences rather than topic choice. The measures in this study provide connecting evidence that creates a big picture of the complex ideological division penetrating public life. The method also surpasses LDA-based approaches for creating common topics. 
 
While most typical use of social media is focused on disease outbreaks, protests, and elections, Khandpur et al. \cite{cyberattack} explored social media to uncover ongoing cyber-attacks. The unsupervised approach detects cyber-attacks such as; breaches of private data, distributed denial of service (DDOS) attacks, and hijacking of accounts while using only a limited set of event trigger as a fixed input.
 
Coordinated campaigns aim to manipulate and influence users on social media platforms. Pacheco et al. \cite{coordcomp} work aim to unravel such campaigns using an unsupervised approach. The method builds a coordination network relying on random behavioural traces shared between accounts. A total of five case studies are presented in work, including U.S. elections, Hong Kong protests, the Syrian civil war, and cryptocurrency manipulation. Networks of coordinated Twitter accounts are discovered in all these cases by inspecting their identities, images, hashtag similarities, retweets, or temporal patterns. The authors propose using the presented approach for uncovering various types of coordinated information warfare scenarios.
 
Coordinated campaigns can also influence people towards offline violence. Xian Ng et al. \cite{coordinariots} investigates the case of capital riots. They introduce a general methodology to discover coordinated by analyzing messages of user parleys on Parler. The method creates a user-to-user coordination network graph prompted by a user-to-text graph and a similarity graph. The text-to-text graph is built on a textual similarity of posts shared on Parler. The study of three prominent user groups in the 6 January 2020 Capitol riots detected networks of coordinated user clusters that posted similar textual content in support of different disinformation narratives connected to the U.S. 2020 elections. 
 
Wanzheng Zhu and Suma Bhat \cite{drgEuphemisticPD} studies the specific case of the use of euphemisms by fringe groups and organizations that is expression substituted for one considered to be too harsh. The work claims to address the issue of Euphemistic Phrase detection without human effort for the first time. Firstly the phrase mining is done on raw text corpus to extract standard phrases; then, word embedding similarity is implemented to select candidates of euphemistic phrases. In the final phases, those candidates are ranked using a masked language model called SpanBERT.
 
Yang Yang et al. \cite{hmntrfk} explore the use of  Network Structure Information (NSI) for detecting human trafficking on social media. They present a novel mathematical optimization framework that combines the network structure into content modelling to tackle the issue. The experimental results are proven effective for detecting information related to human trafficking.

\begin{table}[h!]
  \begin{center}
    \begin{tabular}{p{6cm}c}
    \toprule 
        \textbf{Tweets} & \textbf{Proposed dangerous event type }   \\
         \midrule 
            ``RT @KaitMarieox: This deranged leftist and LGBT activist named Keaton Hill assaulted and threatened to kill @FJtheDeuce, a black conservati…'' & Action-based dangerous event  \\
            ``RT @Lrihendry: When Trump is elected in 2020, I’m outta here. 
It’s a hate-filled sewer. 
It is nearly impossible to watch the hateful at…" &Sentiment-based dangerous event  \\
           ``Scientists predict a tsunami will hit Washington, DC on 1/18/2020
We Are Marching in DC… https://t.co/3af4ZhyV3J" & Scenario-based dangerous event \\
            \bottomrule 
    \end{tabular} \caption{Presumed types of dangerous events for tweets.}     \label{tab:exde}
  \end{center}
\end{table}

Authors present Table \ref{tab:exde} to clarify the intent of this work by providing an example of the collected tweets and their presumed techniques. Based on the existing methods for event detection, it gives a clear objective for using these methods for detecting dangerous events.

\subsection{Event Prediction}
 Event prediction is a complex issue that revolves around many dimensions. Various events are challenging to predict before they become apparent. For example, it is impossible to predict in case of natural disasters, and they can only be detected after the occurrence. Some events can be predicted while they are still in the evolving phase. Authors of \cite{nourbakhsh2017breaking} identify events from local news sources before they may become breaking news globally. The use case of Covid-19 can be regarded as an example where it started locally and became a global issue later. 
 
 A dataset is obtained from a recent Kaggle competition to explore the usability of a method for predicting disaster in tweets. The work in \cite{twitterbertpred} tests the efficiency of BERT embedding, which is an advanced contextual embedding method that constructs different vectors for the same word in various contexts. The result shows that the deep learning model surpasses other typical existing machine learning methods for disaster prediction from tweets. 
 
 Zhou et al. \cite{covidfatlerate} proposed a novel framework called as Social Media enhAnced pandemic suRveillance Technique (SMART) to predict Covid-19 confirmed cases and fatalities. The approach consists of two parts where firstly, heterogeneous knowledge graphs are constructed based on the extracted events. Secondly, a module of time series prediction is constructed for short-and long-term forecasts of the confirmed cases and fatality rate at the state level in the United States and finally discovering risk factors for intervening COVID-19. The approach exhibits an improvement of 7.3\% and 7.4\%  compared to other state-of-the-art methods.
 
 Most of the other existing research targets particular scenarios of event prediction with limited scope. Keeping in mind the complexity of this problem, we only present a few related works, and the generalization is obscure. 
 
\begin{landscape}\centering
\vspace*{\fill}
\begin{table}[htpb]
\begin{tabular}{{llllll}}

 & Event Type & Technique & Reference & Dataset & Year \\ \cline{2-6} 
Scenario-based & & & & & \\ \hline
 & Natural Disasters & SVM/SGD & \cite{nourbakhsh2017breaking} & Twitter & 2017 \\
 & Earthquake & Classification(SVM)& \cite{equake2010} & Twitter & 2010 \\
 & Crisis & CrisisBERT & \cite{crisisbert} &Twitter (C6,C36) & 2021\\
 &Earthquake \&Hurricane & Unsupervised & \cite{hurrearth} &Twitter & 2019 \\
 
 &Fire and Haze Disaster & Classification (hotspots) & \cite{firehaze} &NASA \&Twitter & 2017 \\ 
 & Emergency & Text-CNN, Linear SVC \& Clustering & \cite{emergemcyweibo} & Weibo & 2021 \\
& ... & ... &... &... &...\\ 
 
Sentiment-based & & && & \\ \hline
 & Extreme Sentiments & Unsupervised learning & \cite{extremsentilax} & misc. & 2020 \\
 & Covid19 Sentiments & word2vec & \cite{covsent} & Twitter \& Instagram & 2021 \\
 &Hate speech \& offensive Language& BERT & \cite{hateoffense} & HASOC(Twitter) &2021 \\ 
 &Far-right Extremism & Classification & \cite{farrightextreme} & Facebook, Twitter \&Youtube & 2021 \\ 
 & Political Polarization & Clustering & \cite{plrztnpolitk}& Twitter &2019 \\ 
& ... & ... &... &... &...\\ 
 
Action-based & & & & & \\ \hline
& Cyber attack & Unsupervised & \cite{cyberattack} & Twitter & 2017 \\
& Coordinated campaigns & Unsupervised & \cite{coordcomp} & misc. & 2021 \\ 
& Riots & Clustering & \cite{coordinariots} & Parler & 2021 \\
&Drugs Trafficking& SPANBert & \cite{drgEuphemisticPD} &Text Corpus(subreddit) & 2021\\
& Human Trafficking & Classification (NSI) & \cite{hmntrfk}& Wiebo & 2018 \\
& ... & ... &... &... &...\\ 
\end{tabular}\caption{Dangerous Events categorized under relevant types }\label{detable}
\end{table} 
\vfill
\end{landscape}
 
\section{Conclusion}\label{sec13}
In this work, we laid the basis of the term ``Dangerous Events" and explored different existing techniques and methods for detecting events on social media. Dangerous events contain a broad meaning, but we keep it essential to define the term better. We believe much more can be included in dangerous events, as we explored in the discussion section. Categorizing dangerous events into sub-categories can help specify the event and its features. The subcategories consist of scenario-based, sentiment-based and action-based dangerous events. The usefulness of social media these days provides a significant advantage in detecting such events initially. While in some cases, significant events also originate from social media and manifest in real life, such as; mass protests, communal violence and radicalization.  

The events on social media are mainly polarized. People use it to express their likes or dislike, which can also be classified as positive or negative. Not all extreme events are dangerous, but all dangerous events are extreme, as people can show their happiness using extreme emotions, which is an anomaly under normal circumstances. There is a common element in all dangerous events: we all want to avoid it because of the harm it brings. We believe there is an excellent scope for related work in future. As a proposal, we suggest a dataset containing all types of dangerous events. Secondly, different techniques can be applied to this dataset to further deepen the usefulness and evolve a technique that can be generalized for all kinds of such events. Considering the limitations of event detection, techniques covering only specific events, a joint base can help discover the universally applicable method. 

\backmatter

\bmhead{Acknowledgments}

This work was supported by National Founding from the FCT- Fundacão para a Ciência e a Tecnologia, through the MOVES Project-PTDC/EEI-AUT/28918/2017 and by operation Centro-01-0145-FEDER-000019-C4-Centro de Competências em Cloud Computing, co-financed by the European Regional Development Fund (ERDF) through the Programa Operacional Regional do Centro (Centro 2020), in the scope of the Sistema de Apoio à Investigação Científica e Tecnológica.


\bibliography{sn-bibliography}


\end{document}